\documentclass[10pt,conference]{IEEEtran}
\IEEEoverridecommandlockouts

\usepackage{color}
\usepackage{amsmath}
\usepackage{tcolorbox}
\usepackage{cite}
\usepackage{multirow}
\usepackage{amsmath,amssymb,amsfonts}
\usepackage{algorithmic}
\usepackage{graphicx}
\usepackage{textcomp}
\usepackage{booktabs}
\usepackage{subfigure}
\usepackage{url}
\usepackage{balance}

\newcommand{\mcode}[1]{$\tt #1$}

\newcommand{\apidiff}{{\sc apidiff}}

\def\BibTeX{{\rm B\kern-.05em{\sc i\kern-.025em b}\kern-.08em
    T\kern-.1667em\lower.7ex\hbox{E}\kern-.125emX}}
    
\begin{document}

\title{Why and How Java Developers Break APIs}



\author{\IEEEauthorblockN{Aline Brito\IEEEauthorrefmark{1}, Laerte Xavier\IEEEauthorrefmark{1}, Andre Hora\IEEEauthorrefmark{2}, Marco Tulio Valente\IEEEauthorrefmark{1}}
\IEEEauthorblockA{\IEEEauthorrefmark{1}ASERG Group, Department of Computer Science (DCC), Federal University of Minas Gerais, Brazil\\
\{alinebrito, laertexavier, mtov\}@dcc.ufmg.br
}
\IEEEauthorblockA{\IEEEauthorrefmark{2} Faculty of Computer Science (FACOM), Federal University of Mato Grosso do Sul, Brazil\\
hora@facom.ufms.br}
}

\maketitle

\begin{abstract}
 Modern software development  depends on APIs to reuse code and increase productivity. 
As most software systems, these libraries and frameworks also
 evolve, which may break existing clients.
 However, the main reasons to introduce breaking changes in APIs are unclear. Therefore, in this paper, we report the results 
 of an almost 4-month long field study with the developers of 400
 popular Java libraries and frameworks. We configured an infrastructure
 to observe all changes in these libraries and to detect breaking changes shortly after their introduction in the code. After identifying breaking changes, we   asked the developers to explain the reasons behind their decision to change the APIs.
 During the study, we identified 59 breaking changes, confirmed by the developers of 19
 projects. By analyzing the developers' answers, we report that breaking changes
 are mostly motivated by the need to implement new features,
 by the desire to make the APIs simpler and  with fewer elements, and to improve maintainability. We conclude by providing suggestions to language designers, tool builders, software engineering researchers and API developers.
\end{abstract}

\begin{IEEEkeywords}
API Evolution, Breaking Changes, Field Study.
\end{IEEEkeywords}

\section{Introduction}
\label{section:introduction}

Software libraries are commonly used nowadays to support development, providing source code reuse, improving productivity, and, consequently, decreasing costs~\cite{Mose96, Kons09, raemaekers12}.
For example, there are more than 200K libraries registered on Maven's central repository, a popular package management for Java. They cover distinct scenarios, from mobile and web programming to math and statistical analysis.
These functionalities are provided to client systems via \emph{Application Programming Interfaces} (APIs), which are contracts that clients rely on~\cite{reddy2011api}.
In principle, APIs should be stable and backward-compatible when evolving, so that clients can confidently rely on them.

In practice, however, the literature shows the opposite: APIs are often unstable and backward-incompatible (e.g.,~\cite{Wu10, Robb12, McDo13, thung16}).
A recent study points that 28\% out of 500K API changes break backward compatibility, that is, they may cause side effects on client systems~\cite{laerte17}.
API breaking changes comprise from simple modifications such as the change of a method signature or return type to more critical and dangerous ones such as the removal of a public element. 
In this context, one important question is not completely answered in the literature: \emph{despite being recognized as a programming practice that may harm client applications, why do developers break APIs?}
Better understanding these reasons may support the development of new language features and software engineering approaches and tools to improve library maintenance practices.

In this paper, we study the motivations driving API breaking changes from the perspective of library developers.
By mining daily commits of relevant libraries, we looked for API breaking changes, and, when detected, we sent emails to developers to better understand the reasons behind the changes, the real impact on client applications, and the practices adopted to alleviate the breaking changes.
We also characterize the most common program transformations that lead to breaking changes.
Specifically, we investigate four research questions:

\begin{enumerate}

\item \emph{How often do changes impact clients?} 39\% of the changes investigated in the study may have an impact on clients. However, a minor migration effort is required in most cases, according to the surveyed developers.

\item \emph{Why do developers break APIs?} We identified three major motivations to break APIs, including changes to support new features, to simplify the APIs, and to improve maintainability.

\item \emph{Why don't developers deprecate broken APIs?} Most developers mentioned the increase on maintainability effort as the reason for not deprecating broken APIs.

\item \emph{How do developers document breaking changes?} Most developers plan to document the detected breaking changes, mainly using release notes and changelogs.

\end{enumerate}

By following a firehouse interview method~\cite{diffusionInnovations-2003}, we monitored 400 real world Java libraries and frameworks hosted on GitHub during 116 days.
During this period, we detected 282 possible  breaking changes, sent 102 emails, and received 56 responses, which represents a response rate of
55\%.
With the study, we provide the following contributions: (1) to the best of our knowledge, this is the first large-scale field study that reveals the reasons of concrete breaking changes introduced by practitioners in the source code of popular Java APIs; (2) we  show how breaking changes are introduced in the source code, including the most common program transformations used to break APIs; (3) we provide an extensive list of implications of our study, including implications to language designers, tool builders, software engineering researchers, and practitioners.

\noindent\emph{Structure of the paper.} 
Section~\ref{section:apiDiff} introduces the tool and approach used to detect breaking changes. 
Section~\ref{section:studyDesing} details our experiment design, while Section~\ref{section:results} presents our results.
We discuss the implications of the study in Section~\ref{section:discussion}.
Section~\ref{section:threatsValidity} states threats to validity and Section~\ref{section:relatedWork} presents related work. 
Finally, we conclude the paper in Section~\ref{section:conclusion}.

\section{\apidiff\ Tool}
\label{section:apiDiff}

To detect breaking changes, we use a tool named \apidiff, which was implemented and used by Xavier et al.~\cite{laerte17} in a study about the frequency and impact of breaking changes. 
Essentially, \apidiff\ compares two versions of a library and lists all changes in the signature of public methods, constructors, fields, annotations, and enums. In this paper, the results produced by \apidiff\ are named {\em Breaking Change Candidates} (BCC). The reason is that changes in public elements---as identified by \apidiff---do not necessarily have an impact on API clients. For example, the changed elements may denote internal or low-level services, which are designed only for local usage. To clarify this question, we conducted a survey with API developers, to confirm whether the BCCs detected by \apidiff\ are indeed \emph{breaking changes} (see Section~\ref{section:studyDesing}).

\begin{tcolorbox}[left=0mm,right=0mm,boxrule=0.25mm,colback=gray!5!white]
\vspace{-0.1cm}
{\em Definition:} Changes detected by \apidiff\ in public API elements are named Breaking Change Candidates (BCC).
\vspace{-0.5cm}
\end{tcolorbox}

\begin{figure}[!htp]
	\centering
    \includegraphics[width=0.49\textwidth]{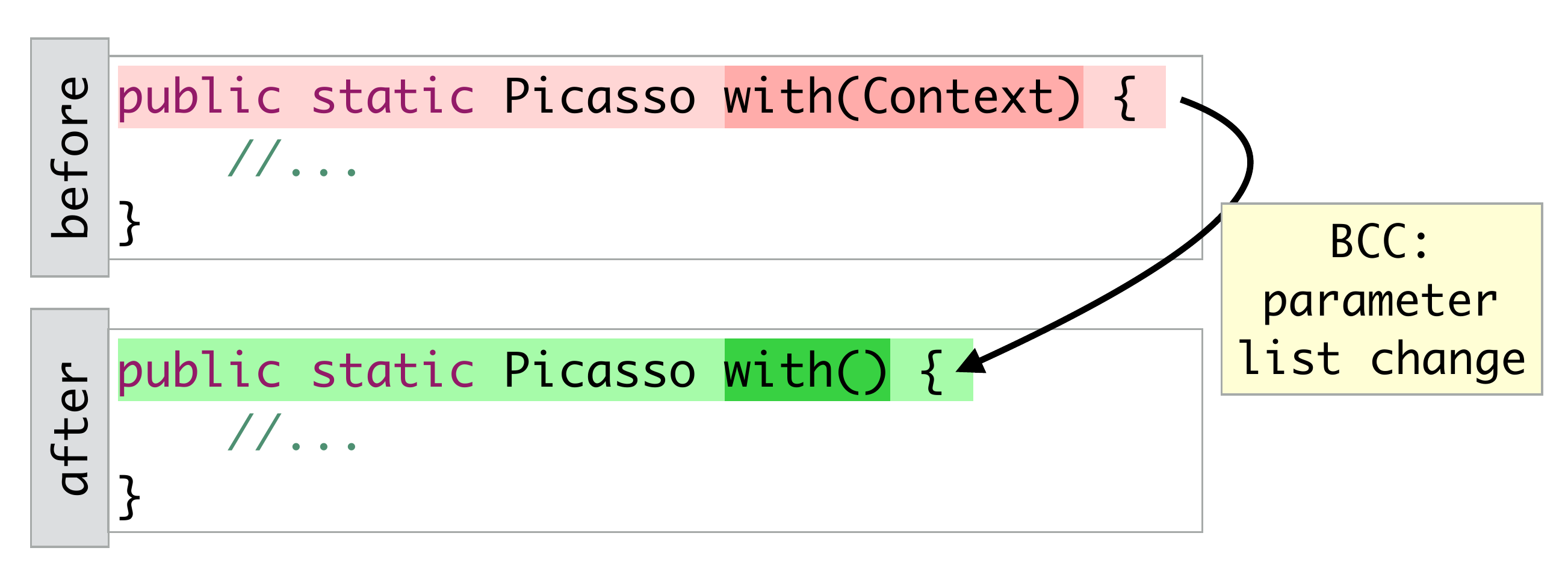}
	\caption{Example of BCC detected by \apidiff~at method level}
	\label{fig:example_bc_apidiff}
\vspace{-0.3cm}    
\end{figure}

Table \ref{table:catalogAPIDiff} lists the BCCs detected by \apidiff. These changes refer to the following API elements: types, methods, or fields. BCCs on types include, for example, drastic changes, like the removal of a type from the code. But subtle changes in public types are also detected, including changing a type visibility from public to another modifier, changing the supertype of a type, adding a \textit{final} modifier to a type (to disable inheritance), or removing the {\em static} modifier of an inner class.
Besides the changes detected to types, BCCs in methods include changes in return types or parameter lists. Changes in fields include, for example, changing the default value of a field.
Figure \ref{fig:example_bc_apidiff} shows an example of BCC detected by \apidiff\ in a method of \textsc{square/picasso} (an image downloading library). According to the developer who performed this change, he removed the parameter \mcode{Context} from method \mcode{with} to simplify the API, since this parameter can be retrieved in other ways.

\begin{table}[!ht]
\vspace{-0.25cm}
\centering
\caption{BCCs detected by \apidiff}
\label{table:catalogAPIDiff}
\begin{tabular}{l p{6.4cm}}
\toprule
{\bf Element} & \multicolumn{1}{c}{\bf BCC}
\\ \midrule

Type & 
\textsc{remove class}, \textsc{change in access modifiers}, \textsc{change in supertype}, \textsc{add final modifier}, \textsc{remove static modifier}                                                \\
	                               
Method & 
\textsc{remove method}, \textsc{change in access modifiers},  \textsc{change in return type},  \textsc{change in parameter list},  \textsc{change in exception list},  \textsc{add final modifier}, \textsc{remove static modifier} 
\\

Field & \textsc{remove field}, \textsc{change in access modifiers},  \textsc{change in field type}, \textsc{change in field default value}, \textsc{add final modifier}     \\
		\bottomrule
\end{tabular}        
\end{table}

As implemented by the current \apidiff\ version, changes in deprecated API elements (i.e.,~elements annotated with \mcode{@Deprecated}) are not BCCs. The rationale is that clients of these elements were previously warned that they are no longer supported, and, therefore, subjected to changes or even to removal. Finally, \apidiff\ warns if a BCC is performed in an experimental or internal API~\cite{businge:2015, mastrangelo:2015ACM}. For this purpose, the tool checks if the qualified name of the changed API element includes a package named \mcode{internal}, as in this example: \mcode{io.reactivex.\underline{internal}.util.ExceptionHelper}. With this warning, the goal is to alert users that the identified BCC is probably a false breaking change.

\begin{figure*}[!t]
\centering
\includegraphics[width=\textwidth]{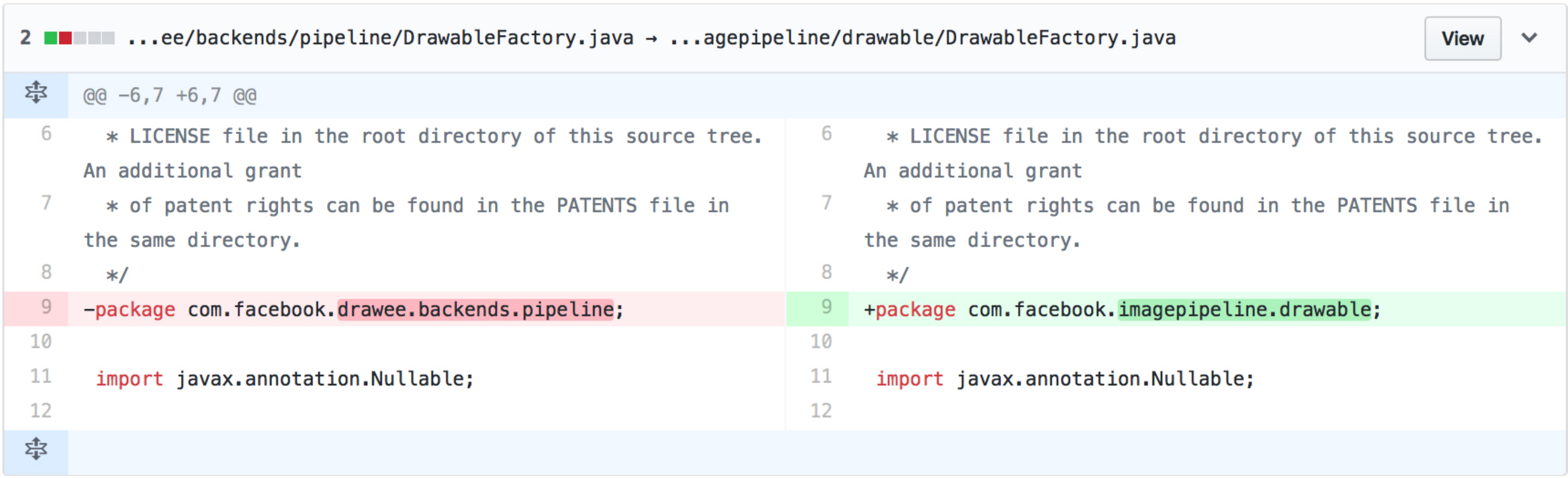}
\caption{Screenshot of a textual diff produced by GitHub in {\sc facebook/fresco}. A {\sc move class} is indicated in the header line.}
\label{fig:diff}
\end{figure*}

As presented in Table~\ref{table:catalogAPIDiff}, \apidiff\ does not use the term refactoring to name BCCs. For example, the {\sc rename} of an API element \mcode{A} to \mcode{B} is identified as the removal of the element \mcode{A} from the code. Similarly, a {\sc move class/method/field} from location \mcode{C} to a new location \mcode{D} is identified as the removal of the element from its original location \mcode{C}. In order to use the most appropriate names to identify these operations, we manually inspected the BCCs detected by \apidiff.
For each commit with a BCC, we analyzed its textual diff, as generated by GitHub. The detection of refactorings performed on classes ({\sc rename/move Class}) was facilitated because these operations are automatically indicated in  the textual diff computed by GitHub. For example, Figure~\ref{fig:diff} shows a
screenshot of a diff in {\sc facebook/fresco} that includes a {\sc move class}.\footnote{\url{https://github.com/facebook/fresco/commit/f6fe6c3}}
At the top of the figure, there is an indication that class \mcode{DrawableFactory} was moved from package \mcode{com.facebook.drawee.backends.pipeline} to package \mcode{com.facebook.imagepipeline.drawable}.
By contrast, to detect {\sc rename/move method/field} we needed to perform a detailed inspection on the diffs results.

\begin{figure*}[!t]
	\centering
    \subfigure{\includegraphics[width=0.24\textwidth]{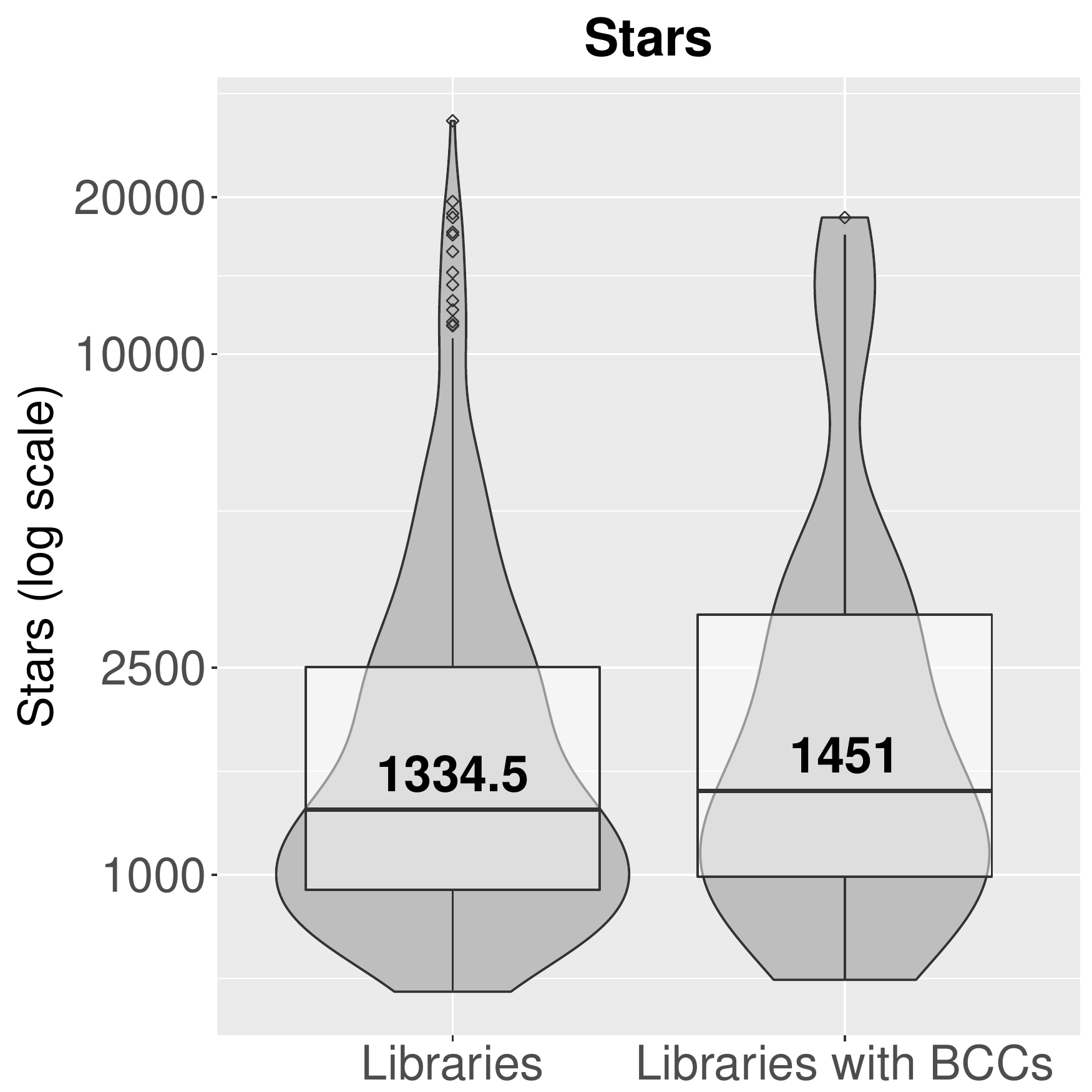}}
	\subfigure{\includegraphics[width=0.24\textwidth]{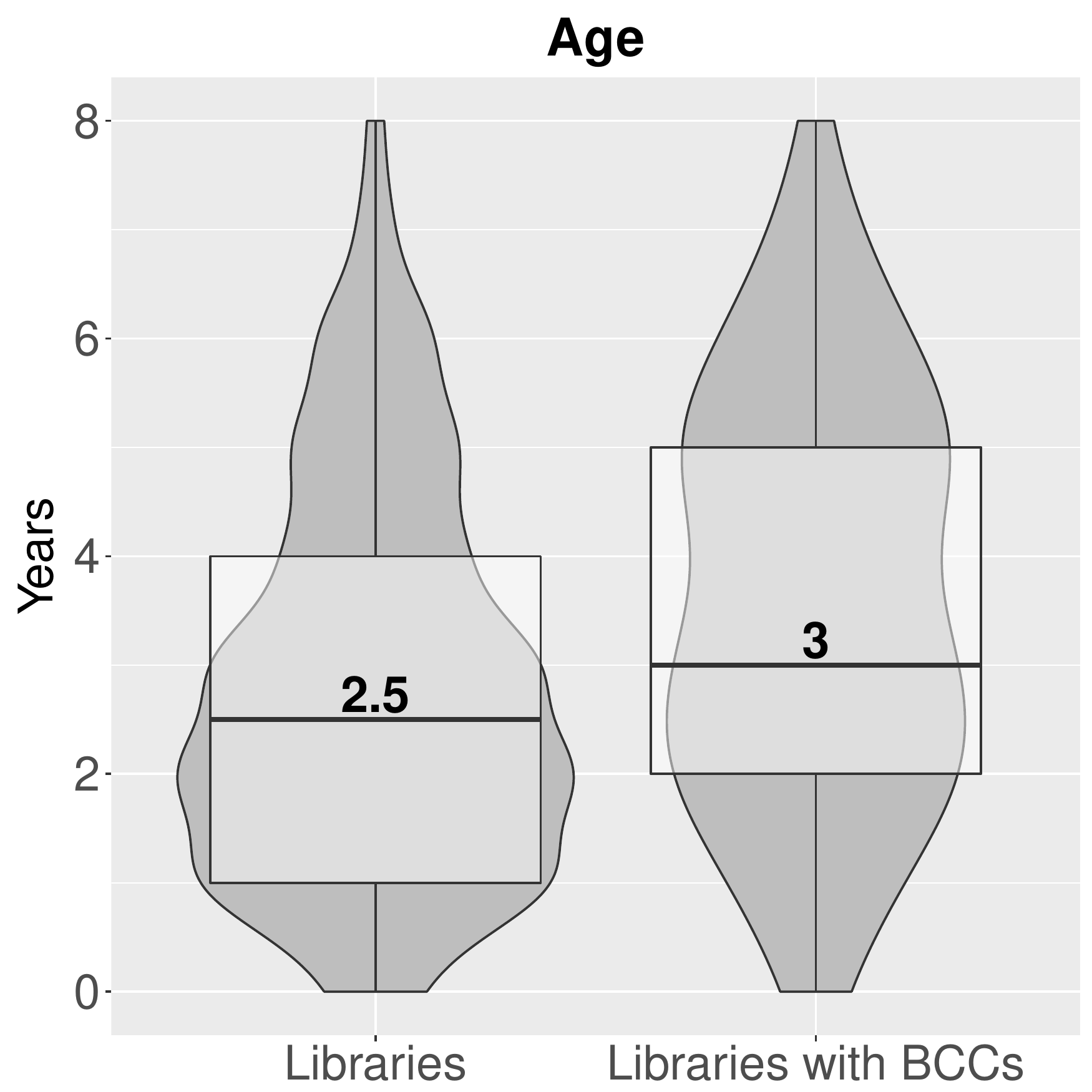}}
	\subfigure{\includegraphics[width=0.24\textwidth]{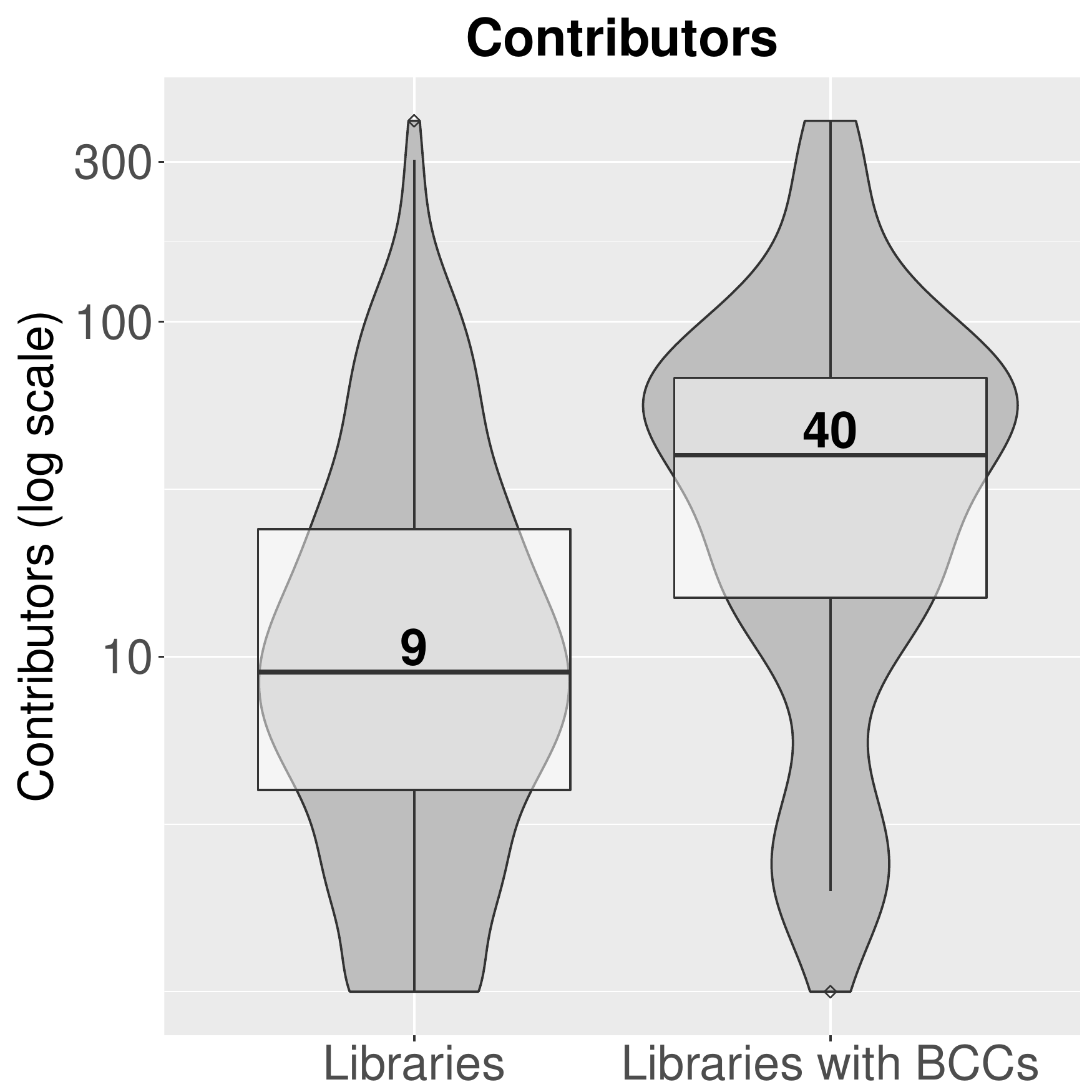}}
	\subfigure{\includegraphics[width=0.24\textwidth]{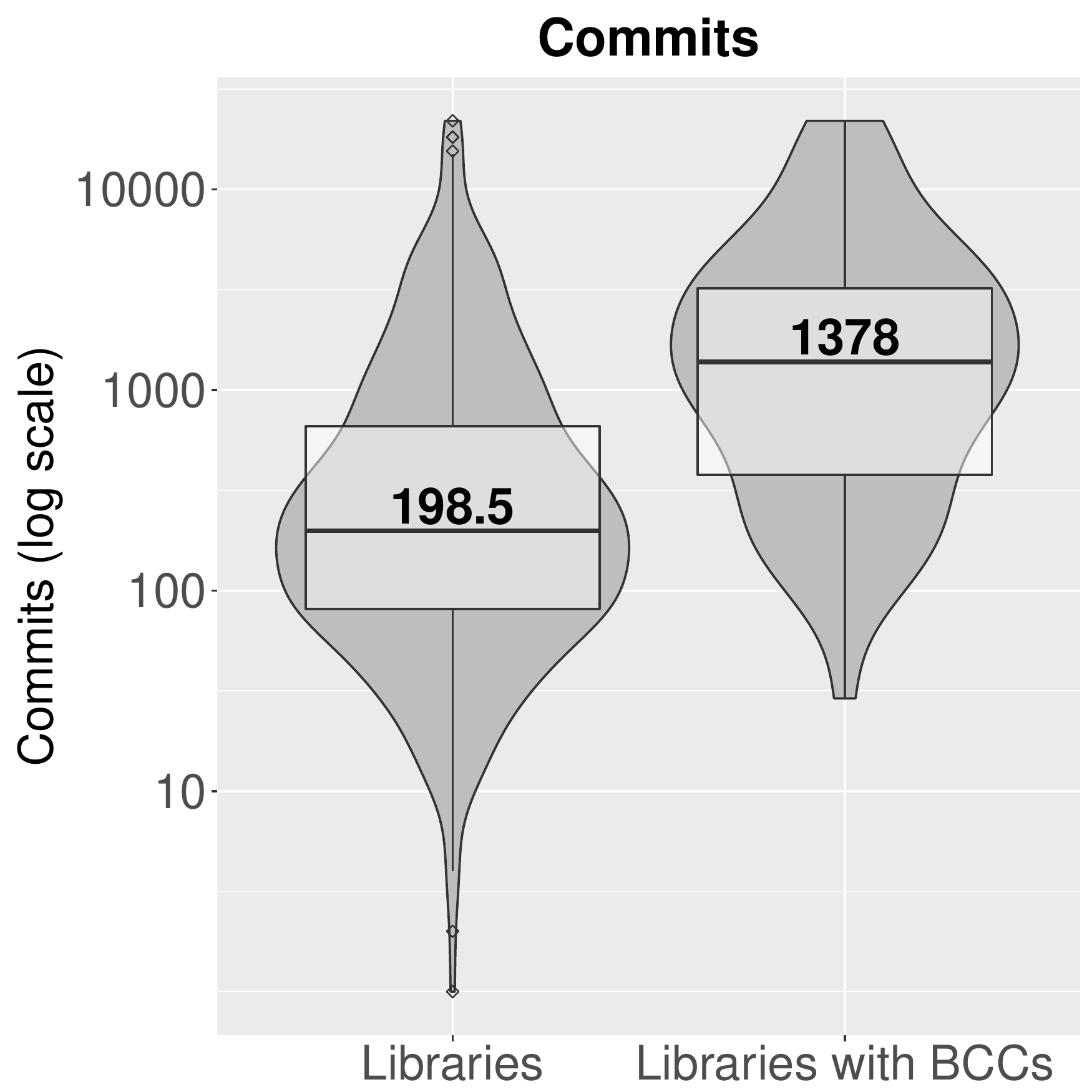}}
	\caption{Distribution of number of stars, age, number of contributors, and number of commits of the initial 400 \emph{Libraries} and of the 61 \emph{Libraries with BCCs}}
	\label{fig:dataset_charac}
\end{figure*}

\section{Study Design}
\label{section:studyDesing}

\subsection{Selection of the Java Libraries}
\label{section:studyDesing-selection-libraries}

First, we selected the top-2,000 most popular Java projects on GitHub, ordered by number of stars and that not are forks (on March, 2017).  We used this criteria because stars is a common and easily accessible proxy for the popularity of GitHub projects \cite{hudson:icsme2016}. Next, we discarded projects that do not have the following keywords in their short description: \textit{library(ies)}, \textit{API(s)}, \textit{framework(s)}.  We also manually removed \textit{deprecated} projects from this list, i.e., projects that have deprecated in their short description, to focus the study on active repositories. These steps resulted in a list of 449 projects. Then, we manually inspected the documentation, wiki, and web pages of these projects to guarantee they are libraries or similar software. As a result, we removed 49 projects. For example, \textsc{googlesamples/android-vision} has the following short description: \textit{Sample code for the Android Mobile Vision API}. Despite having the keyword API in the description, this repository is neither a library nor a framework, but just a tutorial about a specific Android API.
Thus, the final list consists of 400 GitHub projects, including well-known systems such as \textsc{junit-team/junit4} (a testing framework), \textsc{square/picasso} (an image downloading and caching framework), and  \textsc{google/guice} (a dependency injection library).

\subsection{Detecting BCCs}
\label{section:studyDesing-detecting-bcs}

During 116 days, from May 8th to August 31th, 2017, we monitored the commits of the selected projects to detect BCCs. To start the study, on May 8th, 2017  we cloned the selected 400 libraries and frameworks to a local repository. Next, on each work day, we ran scripts that use the \textit{git fetch}
operation to retrieve the new commits of each repository. We discarded a new commit when it did not modify Java files. Furthermore, on Git, developers can work locally in a change and just submit the new revision (via a \textit{git push}) after a while. Therefore, we also discarded commits with more than seven days, to focus the study on recent changes, which is important to increase the chances of receiving feedback from developers (see Section \ref{section:studyDesing-contacting-developers}). We also discarded commits representing merges because these commits usually do not include new features; moreover, merges have two or more parent commits, which leads to a duplication of the BCCs identified by \apidiff~\cite{laerte17, laerte:sanerEra2016}. Finally, we manually discarded commits in branches that only contain test code. 

\begin{figure}[!ht]
\centerline{\includegraphics[width=.3\textwidth]{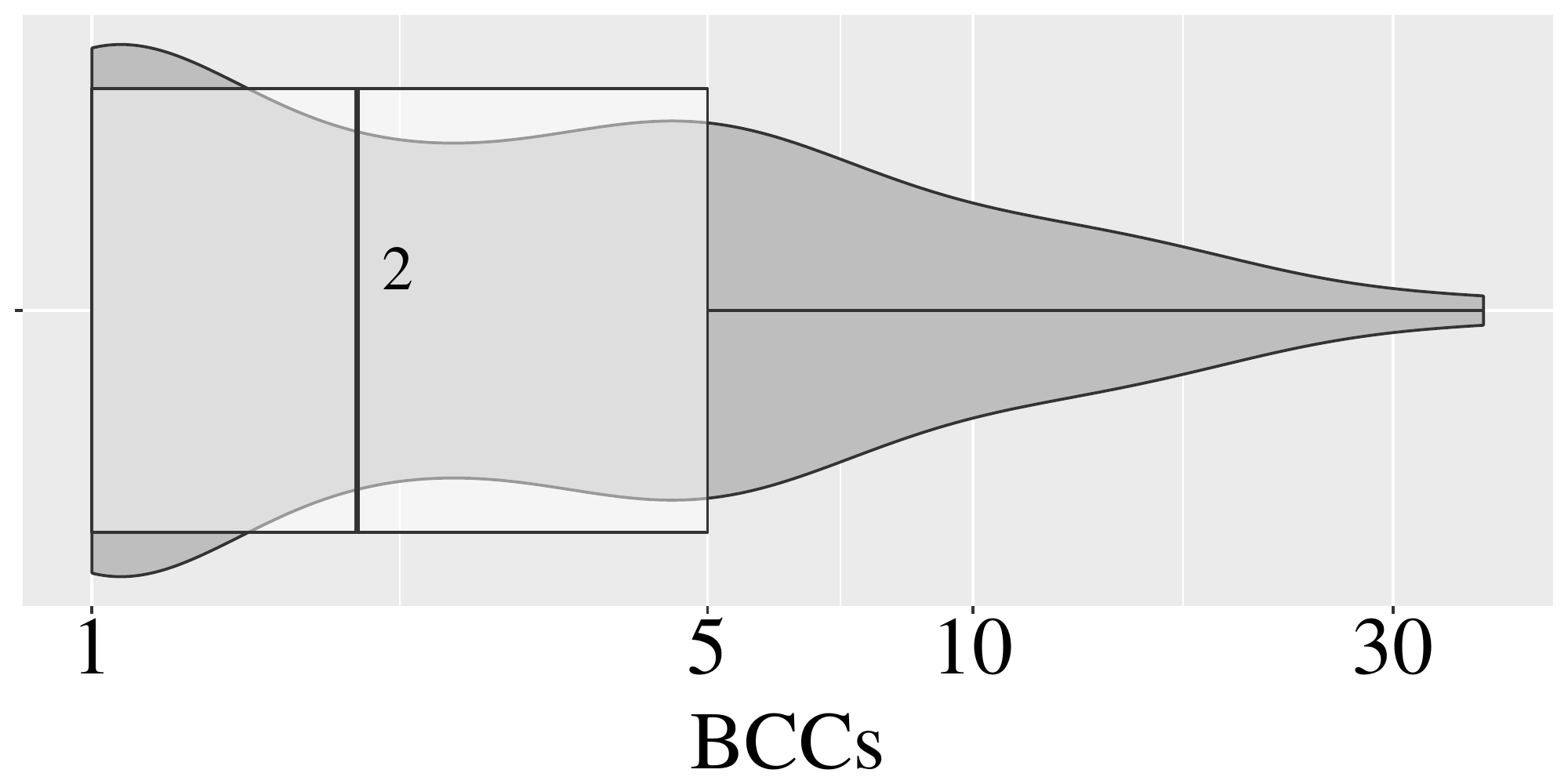}}
\caption{BCCs per project}
\label{fig:bcc-per-project}
\end{figure}

\vspace{-0.1cm}
\apidiff\ identified  282 BCCs in 110 commits, distributed over 61 projects (47\% of the set of 130 libraries and frameworks with commits detected during the study period). Figure~\ref{fig:dataset_charac} presents  the distribution of number of stars, age (in years), number of contributors, and number of commits of the initial selection of 400 libraries and frameworks (labeled as \emph{Libraries}) and of the 61 projects with BCCs (labeled as \emph{Libraries with BCCs}). 
The distributions of \emph{Libraries with BCCs} are statistically different from the initial selection of 400 libraries in age, number of contributors, and number of commits, but not regarding the number of stars (according to Mann-Whitney U Test, $p$-value $\leq$ 5\%). 
To show the effect size of this difference, we computed Cliff's delta (or $d$)~\cite{cliff2014ordinal}.
The effect is medium for age, and large for number of contributors and commits.
In other words, libraries with BCCs are moderately older (3 vs 2.5 years, median measures), but have more contributors (40 vs 9) and more commits (1,378 vs 198.5) than the original list of libraries selected for the study.
Finally, Figure~\ref{fig:bcc-per-project} shows the distribution of BCCs per project, considering only {\em Libraries with BCCs}. The median is two BCCs per project and the system with the highest number of BCCs is {\sc robolectric/robolectric}, with 38 BCCs (including 35 BCCs where public API elements were changed to protected visibility).

\subsection{Contacting the Developers}
\label{section:studyDesing-contacting-developers}

Among the 282 BCCs considered in the study, 268 (95\%) were detected in commits that contain a public email. Therefore, on each day of the study, after detecting such BCCs, we contacted the respective developers. In the emails sent to them (see a template in Figure \ref{fig:templateEmail}), we added a link to the GitHub commit and a description of the BCC. Then, we asked four questions. With the first question, we intended to shed light on the real motivation behind the detected changes. With the second question, we intended to confirm whether the BCC detected by \apidiff\ can break existing clients. With the third question, our interest was to understand why the developers have not deprecated the API element where the BCC was detected. Finally, with the last question, our interest was to investigate how often developers document BCCs.

\begin{figure}[!htp]
\vspace{-0.25cm}
\footnotesize
\begin{tcolorbox}
		
		Dear [developer name],\\
		
		I am a researcher working with API usability and evolution. In my research, I am 	studying the API of [repository/project].\\
		
		I found that you performed the following changes in this project:\\
		
		[BCCs list] and [commit links]\\
		
		Could you please answer the following questions:\\
		
		1.  Why did you perform these changes?\\\\
		2.  Do you agree these changes can break clients? If yes, could you quantify the amount of work to use the new implementation?\\\\
		3.  Why didn't you deprecate the old implementation?\\\\
		4.  Do you plan to document the changes? If yes, how?
\end{tcolorbox}
\caption{Mail to the authors of commits with BCCs detected by \apidiff}
\label{fig:templateEmail}
\end{figure}
		
We sent only one email to each developer. Specifically, whenever we detected BCCs by the same developer, but in different commits, we only sent one email to him, about the BCC detected in the first commit. In this way, we reduced the chances that developers perceived our emails as spam. 
It is also important to mention that before sending each email we inspected the respective commit description to guarantee it did not include an answer to the proposed questions.  In the case of  six commits, we found answers to the first question (\textit{why did you perform these changes?}). As an example, we have the following commit description: \\[-0.2cm]

\noindent{\em
\textit{Lock down assorted APIs that aren't meant to be used publicly subtyped.} (D23, Add Final Modifier)
}\\[-0.2cm]   

In this message, the developer mentions he is adding a \mcode{final} modifier to  classes that must not be extended by API clients. We also sent a brief email to the authors of these six commits, just asking them to confirm that the detected BCCs can break existing clients;
we received two positive answers. Finally, in two commits we found a message describing the motivation for the change and confirming that it is a breaking change. As an example, we have this answer:\\[-0.2cm]

\noindent{\em \textit{Now, [Class Name]  can be configured to apply to different use cases \ldots\ Breaking changes:  Remove [Class Name]} (D22)
}\\[-0.2cm]   

During the 116 days of the study, we sent 102 emails and received 56 responses, which represents a response ratio of 55\%. Table~\ref{table:study-design-numbers} summarizes the numbers and statistics about the study design phase, as previously described in this section. After receiving all emails, we analyzed the answers using thematic analysis \cite{cruzes:esem2011}, a technique for identifying and recording {\em themes} (i.e.,~patterns) in textual documents. Thematic analysis involves the following steps: (1) initial reading of the answers, (2) generating a first code for each answer, (3) searching for themes among the proposed codes, (4) reviewing the themes to find opportunities for merging, and (5) defining and naming the final themes. Steps 1 to 4 were performed independently by two authors of this paper. After this, a sequence of meetings was held to resolve conflicts and to assign the final themes (step 5).
When quoting the answers, we use labels D1 to D60 to indicate the respondents (including four developers with answers coming from commits). 

\begin{table}[!ht]
\vspace{-0.25cm}
\centering
\small
\caption{Numbers about the study design}
\label{table:study-design-numbers}
\begin{tabular}{l r}
\toprule
Days & 116 \\
Projects & 400 \\
Projects with commits & 130 \\
Projects with commits and BCCs & 61 \\
BCCs detected by \apidiff & 282 \\
BCCs in commits with public emails  & 268 \\
Commits confirming/describing BCCs motivations & 4 \\ 
Emails sent to authors of commits with BCCs & 102 \\
Received answers & 56 \\
Response ratio & 55\% \\
\bottomrule
\vspace{-0.5cm}
\end{tabular}        
\end{table}

\section{Results}
\label{section:results}

\subsection{How Often do Changes Impact Clients?}
\label{section:results-unconfirmed-bcs}

To answer this question, we first define breaking changes:
\begin{tcolorbox}
[left=0mm,right=0mm,boxrule=0.25mm,
colback=gray!5!white]
\vspace{-0.1cm}
{\em Definition:} BCCs confirmed by the surveyed developers are named Breaking Changes (BC).
\vspace{-0.2cm}
\end{tcolorbox}

\begin{figure}[!ht]
\centerline{\includegraphics[width=.5\textwidth]{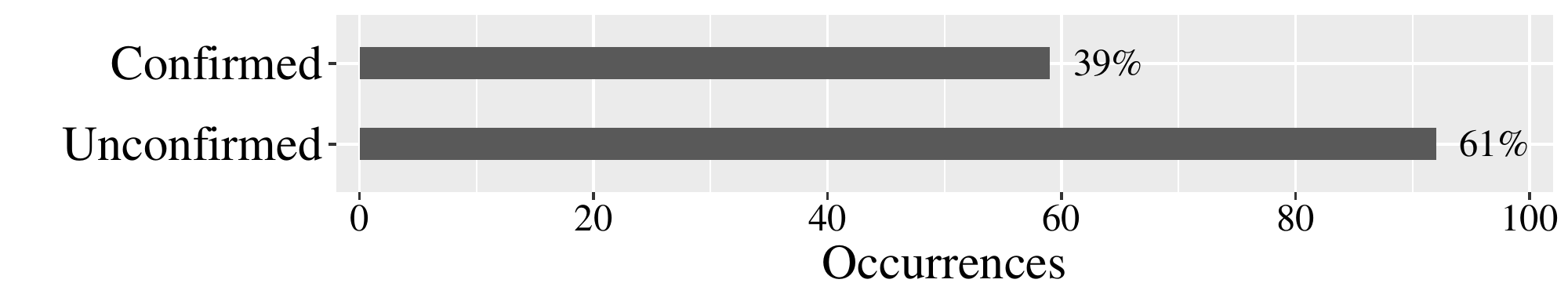}}
	\caption{Confirmed and unconfirmed BCCs; confirmed BCCs are called BCs}
	\label{fig:plot-bcs-and-nonbc}
\end{figure}

As presented in Figure \ref{fig:plot-bcs-and-nonbc}, only 59 BCCs (39\%) detected by \apidiff\ are BCs. The remaining BCCs---which have not been confirmed by the respective developers---are called {\em unconfirmed BCCs}. Next, we characterize the BCs investigated in this study; we also reveal the reasons for the high percentage of unconfirmed BCs.\\[-0.2cm]

\begin{figure}[!tp]
\centerline{\includegraphics[width=.5\textwidth]{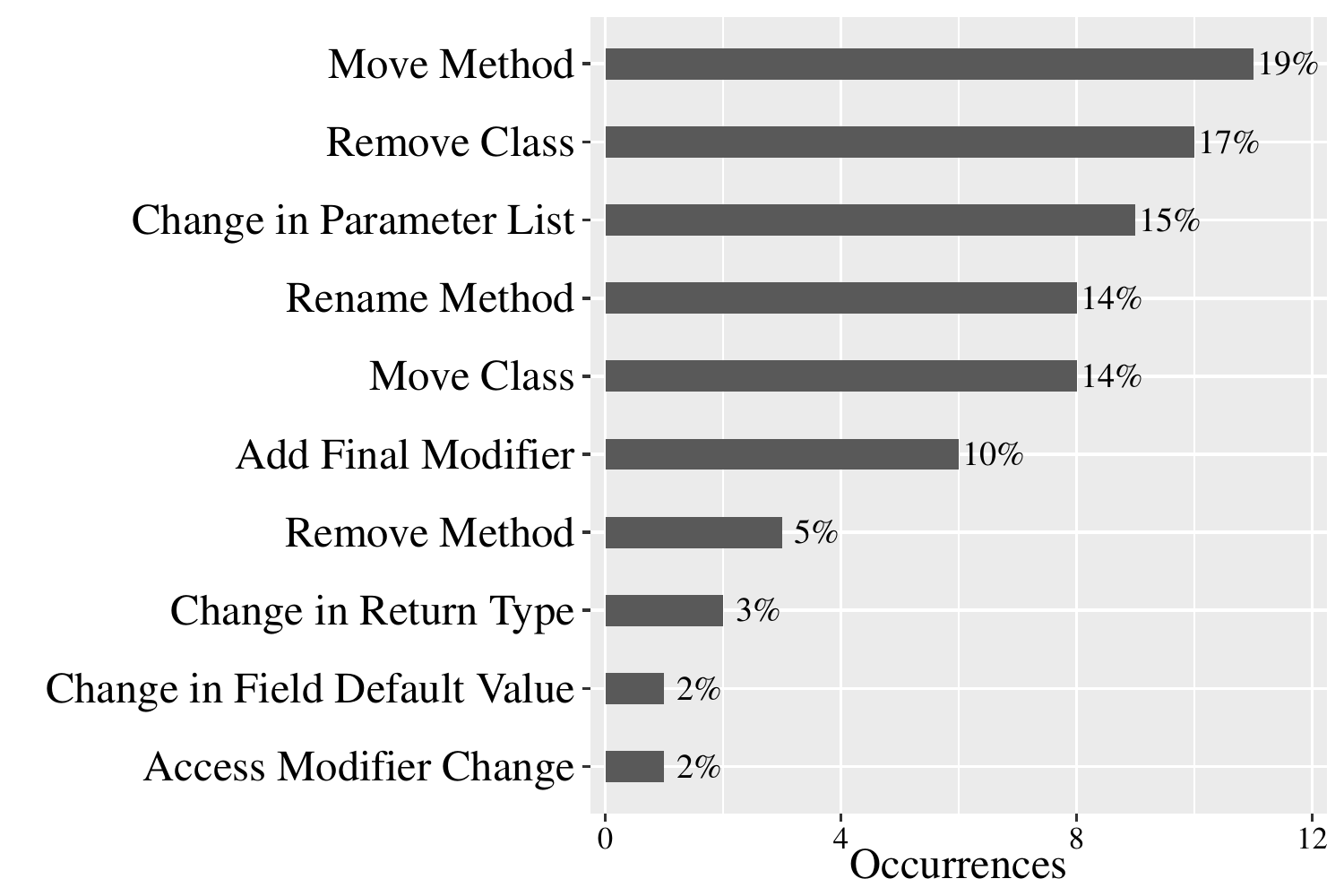}}
\caption{Most common breaking changes}
\label{fig:plot-commom-bcs}
\end{figure}

\begin{figure}[!tp]
\centerline{\includegraphics[width=.5\textwidth]{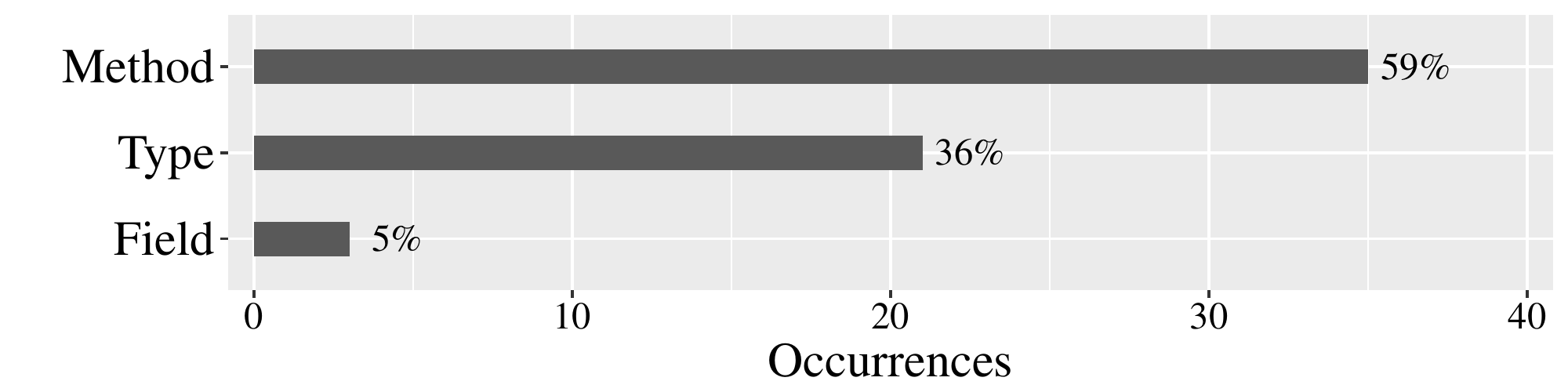}}
\caption{Most common breaking changes per API element}
\label{fig:plot-commom-bcs-per-element}
\end{figure}

\vspace{-0.5cm}
\noindent{\em Breaking Changes (BC):} The 59 BCs detected in the study are distributed over 19 projects and 24 commits, including 20 commits with BCs confirmed by email and 4 commits with BCs declared in the commit description. Figure \ref{fig:plot-commom-bcs} shows the most common BCs. Among the Top-5, three are refactorings,
including {\sc move method} (11 occurrences),
{\sc rename method} (8 occurrences), and
{\sc move class} (8 occurrences). The second most common BCs are the removal of an entire class (10 occurrences), which can be viewed as a drastic API change. The third most popular BCs are {\sc changes in method parameters} (9 occurrences). 
Considering the 17 types of BCs detected by \apidiff\ (see Table~\ref{table:catalogAPIDiff}), only 8 appeared in our study. Regarding the elements affected by the changes, Figure \ref{fig:plot-commom-bcs-per-element} shows the BCs grouped by API element: 35 BCs (59\%) are performed on methods, followed by BCs on types (21 instances, 36\%) and fields (3 instances, 5\%).\\[-0.2cm]

\begin{tcolorbox}[left=0mm,right=0mm,boxrule=0.25mm,colback=gray!5!white]
\vspace{-0.2cm}
{\em Summary:} The most common BCs are due to refactorings (47\%); most BCs are performed on methods (59\%).
\vspace{-0.2cm}
\end{tcolorbox}

\noindent{\em Unconfirmed BCCs:} By contrast, in the case of 92 changes (61\%), the surveyed developers did not agree they have an impact on clients.
We organized the reasons mentioned by these developers on two major themes: internal APIs and testing branches/new releases.
Regarding the first theme, \apidiff\ gives a warning about APIs that are likely to be internal; specifically, the ones implemented in packages containing the string \mcode{internal}, as recommended in the related literature~\cite{businge:2015, mastrangelo:2015ACM}. Nonetheless, 21 developers mentioned that the BCCs occurred at internal (or low-level) APIs that do not include \mcode{internal} in their names,
as in the following answers:\\[-0.2cm]

\noindent{\em
This method is used internally, though it was public. We don't expect people using this method in their applications. (D30)
}\\[-0.2cm]

\noindent{\em
This could potentially break but this class is used internally as utility and not intended to be used by library users. (D32)
}\\[-0.2cm]

The second cause of unconfirmed BCCs are due to testing branches. As described in Section \ref{section:studyDesing-detecting-bcs},  we monitored all branches of the analyzed repositories to contact the developers just after the changes. Consequently, in some cases, we considered BCCs in branches that do not represent major developments, e.g.,~testing branches, branches dedicated to experiments, etc. Ten developers mentioned that the BCC occurred in such branches, as in the following answer:\\[-0.2cm]

\noindent{\em
{This is a early extension of [Project Name] to support Java 9 modules. Thus, the code is neither stable nor complete.} (D42)
}\\[-0.2cm]
\begin{tcolorbox}[left=0mm,right=0mm,boxrule=0.25mm,colback=gray!5!white]
\vspace{-0.1cm}
{\em Summary:} Most unconfirmed BCCs are related to changes in internal or low-level APIs or in testing branches.
\vspace{-0.2cm}
\end{tcolorbox}

\subsection{Why do Developers Break APIs?}
\label{section:results-why-break-apis}

As reported in Table~\ref{table:motivationsBC}, we found four  distinct reasons for breaking APIs: New Feature, API Simplification, Improve Maintainability, and Bug Fixing. In the following paragraphs, we describe and give examples of each of these motivations.

\begin{table}[!h]
\vspace{-0.3cm}
	\centering
	\caption{Why do we break APIs?}
	\label{table:motivationsBC}
	\begin{tabular}{lp{4.4cm}r}
		\toprule
		\multicolumn{1}{c}{\textbf{Motivation}} & \multicolumn{1}{c}{\textbf{Description}}
     & \multicolumn{1}{c}{\textbf{Occur.}} \\ \midrule
		\textsc{new feature}                  & BCs to implement new features                                                        & 19                                       \\
		\textsc{api simplification}               & BCs to simplify and reduce the API complexity and number of elements                          & 17                                       \\
\textsc{Maintainability } & BCs to improve the maintainability and the structure of the code & 14  \\

\textsc{bug fixing}                              & BCs to fix bugs in the  code                                  & 3                                        \\
      	\textsc{other}                      				 & BCs not  fitting the previous cases                                         & 6                                        
\\ 	\bottomrule

	\end{tabular}
\end{table}

\noindent {\em New Feature.} With 19 instances (32\%), the implementation
 of a new feature is the most common motivation to break APIs.  As examples,
 we have the following answers:\\[-0.2cm]

\noindent{\em
The changes in this commit were just a setup before implementing a new feature: chart data retrieval. (D01)
}\\[-0.2cm]

\noindent{\em
The changes were adding new functionality, which were requested on GitHub by the users, but to avoid unnecessary duplications I had to change the method name to better reflect what the method would be doing after the changes. (D13)
}\\[-0.2cm]

In the first answer, D01 moved some classes from packages, before starting the implementation of a new feature. Therefore, clients should update their \mcode{import} statements, to refer to the new class locations. In the second answer, D13 renamed a method to better reflect its purpose after implementing a new feature.  The rename should then be propagated to the method calls in the API clients. 
\\[-0.2cm]

\noindent {\em API Simplification.} With 17 instances (29\%), these BCs include the removal of API elements, to make the API simpler to use. As examples, we have these answers:\\[-0.2cm]

\noindent{\em
We can access the argument without it being provided using another technique. (D03, Change in Parameter List)
}\\[-0.2cm]

\noindent{\em
This method should not accept any parameters, because they are ignored by server. (D08, Change in Parameter List)
}\\[-0.2cm]

\noindent{\em
We are preparing for a new major release and cleaning up the code aggressively. (D09, Remove Class)
}\\[-0.2cm]

In the first two answers, D03 and D08 removed one parameter from public API methods. In the third answer, D09 removed a whole class from the API, before moving to a new major release. In these three examples, the API became simpler and easier to use or understand. However, existing clients must adapt their code to benefit from these changes.\\[-0.2cm]

\noindent {\em Improve Maintainability.} With 14 instances (24\%), BCs performed to improve maintainability, i.e.,~internal software quality aspects, are the third most frequent ones. As examples, we have the following answers:\\[-0.2cm]

\noindent{\em
Because the old method name contained a typo. (D15, Rename Method)
}\\[-0.2cm]

\noindent{\em
Make support class lighter, by moving methods to Class and Method info. (D24, Move Method)
}\\[-0.2cm]

In the first answer, D15 renamed a method to fix a spelling error, while in the second answer, D24 moved some methods to a utility class to make the master class lighter.\\[-0.2cm]

\noindent {\em Bug Fixing.}  In the case of 3 BCs (5\%), the motivation is related with fixing a bug, as in the following answers:\\[-0.2cm]

\noindent{\em
The iterator() method makes no sense for the cache. We can not be sure that what we are iterating is the right collection of elements. (D05, Remove Method)
}\\[-0.2cm]

\noindent{\em
The API element could cause serious memory leaks. (D12, Change in Parameter List)
}\\[-0.2cm]

In the first answer, D05 removed a method with an unpredicted behavior in some cases. In the second answer, D12 removed a flag parameter related to memory leaks.\\[-0.2cm]

\noindent {\em Other Motivations.} This  category includes six BCs whose motivations do not fit the previous cases. For example, BCs performed to remove deprecated dependencies (2 instances), BCs to adapt to changes in requirements and specification (2 instances), BCs to eliminate trademark conflicts (1 instance), and one BC with an unclear motivation, i.e.,~we could not understand the specific answer provided by the developer.

\begin{tcolorbox}[left=0mm,right=0mm,boxrule=0.25mm,colback=gray!5!white]
\vspace{-0.2cm}
{\em Summary:} BCs are mainly motivated by the need to implement new features (32\%), to simplify the API (29\%), and to improve maintainability (24\%).
\vspace{-0.2cm}
\end{tcolorbox}

\begin{figure}[!t]
\centering

\subfigure[fig:top-3-bcs-b][BCs to  implement new features]{\includegraphics[width=0.47\textwidth]{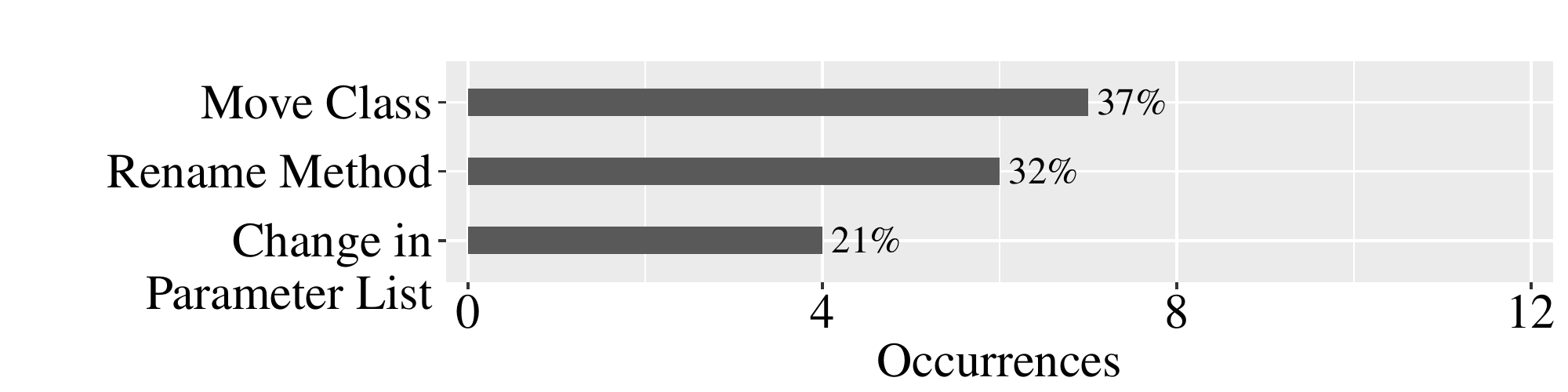}
}

\vspace{-0.3cm}

\subfigure[fig:top-3-bcs][BCs to simplify APIs]{\includegraphics[width=0.47\textwidth]{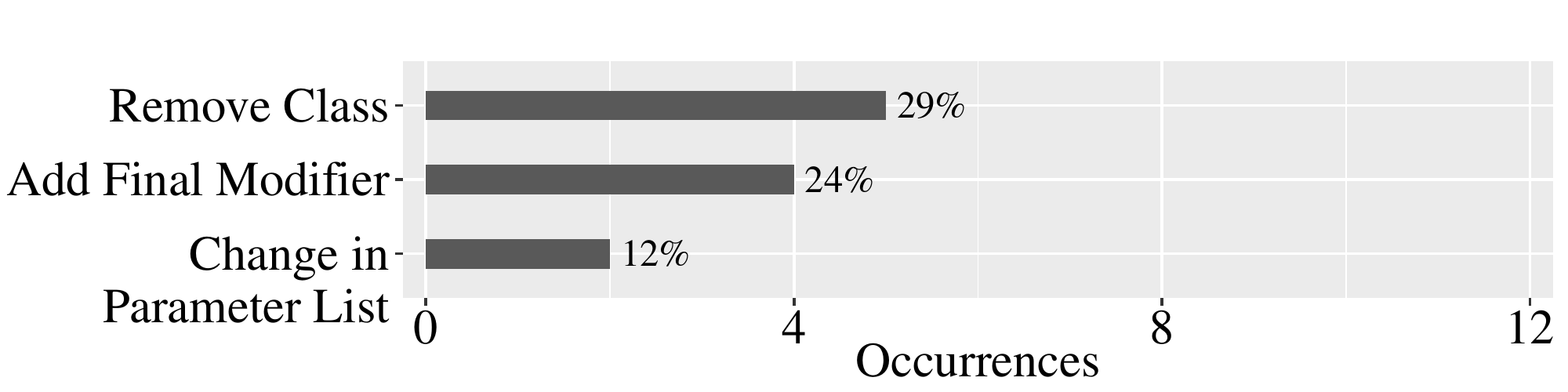}
}	

\vspace{-0.3cm}

\subfigure[fig:top-3-bcs-a][BCs to improve maintainability]{
\includegraphics[width=0.47\textwidth]{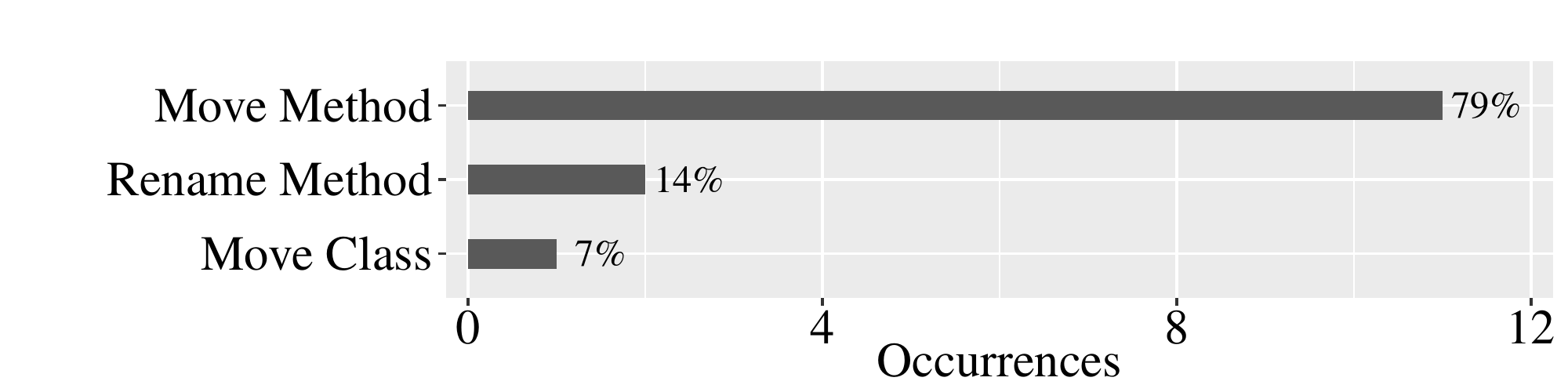}
}

\caption{Top-3 most common BCs,  grouped by motivation}
\label{fig:top3-per-motivation}
\end{figure}

Figure~\ref{fig:top3-per-motivation} shows the top-3 most common BCs due to Feature Addition, API Simplification, and to Improve Maintainability. {\sc move class} is the most common BC when implementing a new feature, with 7 occurrences. Specifically, when working on a new major release, developers tend to start by performing structural changes in the code, which include moving classes between packages. To simplify APIs, developers usually {\sc remove classes} (5 instances) and also add a \mcode{final} modifier to methods (4 instances). The latter is considered a simplification because it restricts the usage of API methods; after the change, the  API methods cannot be redefined in subclasses, but only invoked by clients.
Finally, it is not a surprise that BCs performed to improve maintainability are refactorings. In this case, the three most popular BCs are due to {\sc move method} (11 instances),
{\sc rename method} (2 instances), and {\sc move class} (1 instance). Interestingly, {\sc move class} is 
also used when implementing a new feature.\\[-0.2cm]  

\begin{tcolorbox}[left=0mm,right=0mm,boxrule=0.25mm,colback=gray!5!white]
\vspace{-0.2cm}
{\em Summary:} BCs due to refactorings are performed both to improve maintainability and to enable and facilitate the implementation of new features.
\vspace{-0.2cm}
\end{tcolorbox}

\subsection{What Is the Effort on Clients to Migrate?}
\label{section:results-effort-clients}

\begin{figure}[!tp]
\centerline{\includegraphics[width=.5\textwidth]{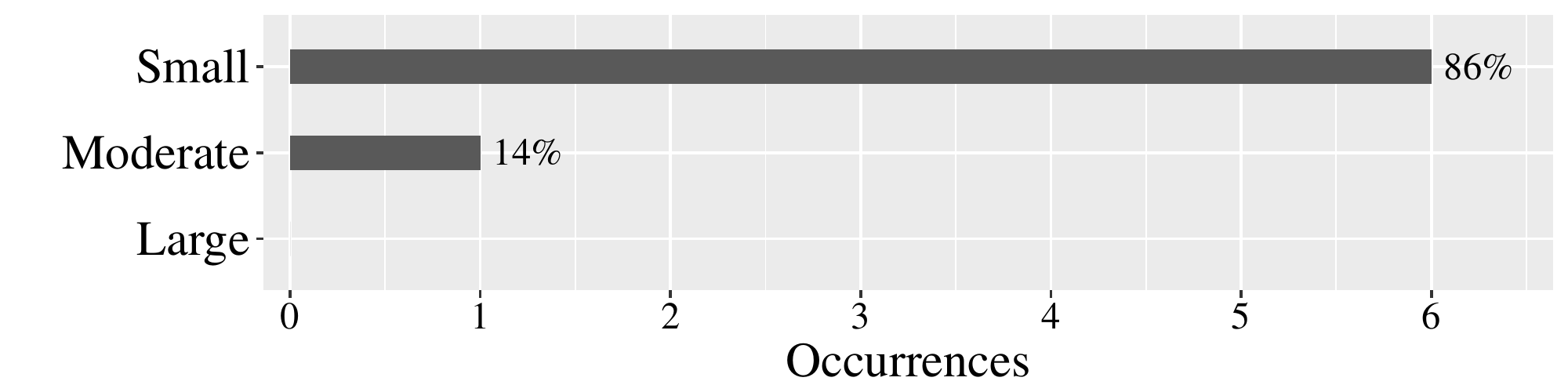}}
\caption{Effort required on clients to migrate}
\label{fig:plot-effort}
\end{figure}

We organized the answers of this survey question in three levels: \textit{minor}, \textit{moderate}, or \textit{major effort}. Seven developers answered the question. As presented in Figure \ref{fig:plot-effort}, six developers estimated that the effort to use the new version is minor, while one answered with a \textit{moderate} effort; none of them considered the update effort as a {\em major} one.
For example, developer D04---who moved a class between packages with the purpose of improving maintainability ---estimates a minor effort on clients to use the new version:\\[-0.2cm]

\noindent{\em
Work required should be minor, since it is just a change of a package. (D04, Move Class)}\\[-0.2cm]

\begin{figure}[!tp]
\vspace{-0.3cm}
\centerline{\includegraphics[width=.5\textwidth]{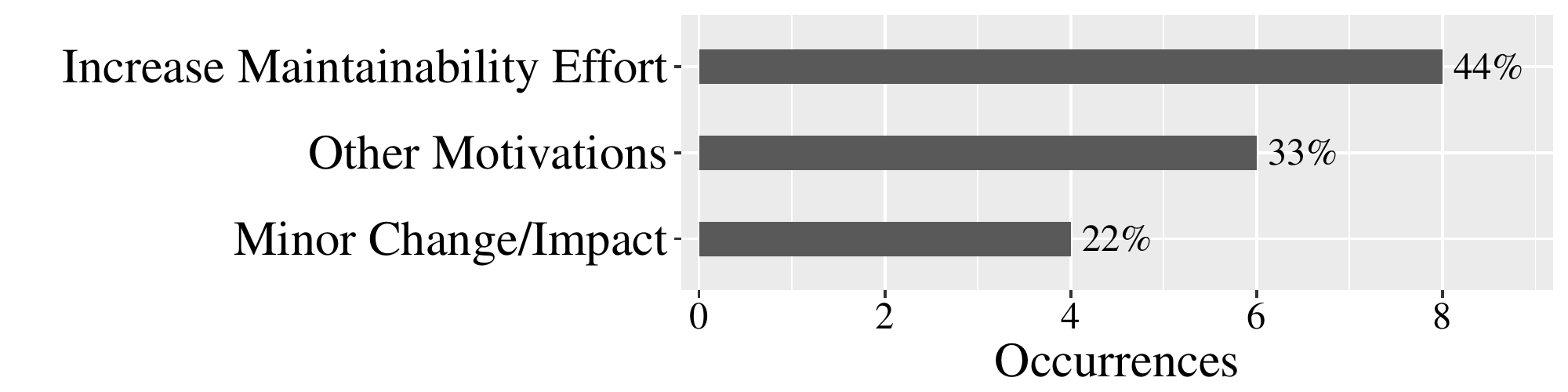}}
\caption{Reasons for not deprecating the old versions}
\label{fig:plot-deprecate}
\vspace{-0.3cm}
\end{figure}

A single developer (D09) answered  that a class removal may require a {\em moderate} effort on clients:\\[-0.2cm]

\noindent{\em The complexity will depend largely on the size of the project and how they use the library. (D09, Remove Class)
}\\

\begin{tcolorbox}[left=0mm,right=0mm,boxrule=0.25mm,colback=gray!5!white]
\vspace{-0.2cm}
{\em Summary:} According to the surveyed developers, the effort on clients to migrate to the new API versions is minor.
\vspace{-0.2cm}
\end{tcolorbox}

\subsection{Why didn't you Deprecate the Old Implementation?}
\label{section:results-deprecate}

17 developers answered this survey question. As presented in Figure \ref{fig:plot-deprecate}, they presented five reasons for not deprecating the API elements impacted by the BCs.\\[-0.2cm]

\noindent {\em Increase Maintenance Effort.} 8 developers mentioned that deprecated elements increase the effort to maintain the project, as in the following answer:\\[-0.2cm]

\noindent{\em
{In such a small library, deprecation will only add complexity and maintenance issues} in the long run. (D16)
}\\[-0.2cm]

\noindent {\em Minor Change/Impact:} Four developers argued that the performed BCs require trivial changes on clients or that the library has few clients, as in the following answers:\\[-0.2cm]

\noindent{\em
Because the fix is so easy. (D15)
}\\[-0.2cm]

\noindent{\em
The main reason is that [the number of] users is small. (D14)
}\\[-0.2cm]

Other motivations include the following ones: 
library is still in beta (1 developers), incompatible dependencies with the old version (1 answer), and trademark conflicts (1 answer). Finally, one developer forgot to add deprecated annotations.\\[-0.2cm]

\begin{tcolorbox}[left=0mm,right=0mm,boxrule=0.25mm,colback=gray!5!white]
\vspace{-0.1cm}
{\em Summary:} Developers do not deprecate elements affected by BCs mostly due to the extra effort to maintain them.
\vspace{-0.2cm}
\end{tcolorbox}

\subsection{How do Developers Document Breaking Changes?}
\label{section:results-document}

\begin{figure}[!htp]
\centerline{\includegraphics[width=.5\textwidth]{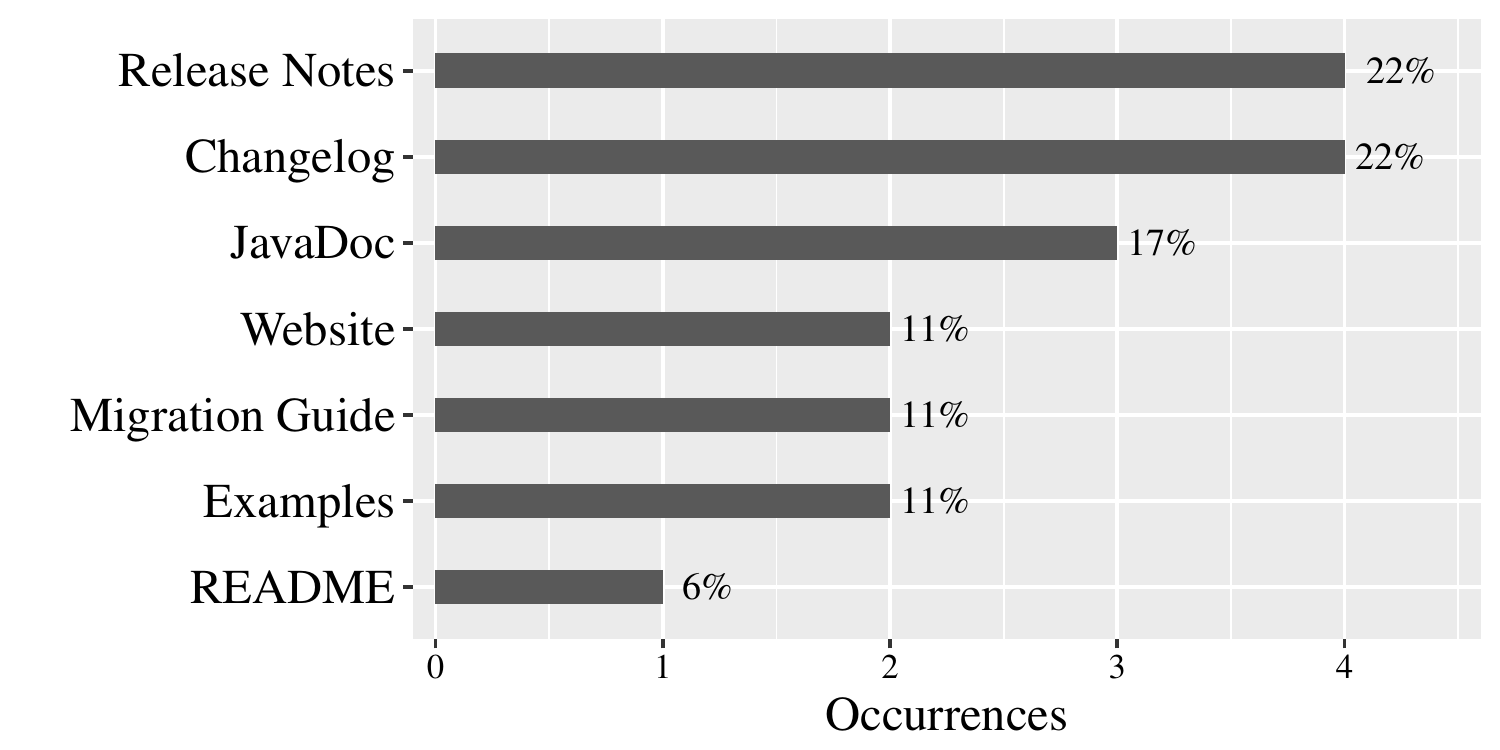}}
\caption{How do you plan to document the detected BCs?}
\label{fig:document-changes}
\end{figure}

This question was answered by 18 developers. Among the received answers, 14 developers stated they intend to document the BCs. We analyzed these answers and extracted seven different documents they plan to use to this purpose (see Figure \ref{fig:document-changes}). Release Notes  and Changelogs are the most common documents, mentioned by four developers each, followed by  JavaDoc (3 developers).

Finally, four developers do not  plan to document the BCs. For example, two of them considered the changes trivial and self-explained.\\[-0.2cm]



\begin{tcolorbox}[left=0mm,right=0mm,boxrule=0.25mm,colback=gray!5!white]
\vspace{-0.2cm}
{\em Summary:} BCs are usually documented using release notes or changelogs. 
\vspace{-0.2cm}
\end{tcolorbox}

\section{Implications}
\label{section:discussion}

This section presents the study implications to language designers, tool builders, researchers, and practitioners.
\\[-0.2cm] 

\noindent{\em Language Designers:} Among the 151 Breaking Changes Candidates (BCCs) with developers' answers, only 59 were classified as Breaking Changes (BCs). The other BCCs are mostly changes in internal or low-level APIs or changes performed in experimental branches. Since they are designed for internal usage only, developers do not view changes in these APIs as BCs. However, previous research has shown that occasionally internal APIs are used by external clients~\cite{andre:fse2016, boulanger-ManagingConcern-2006-icsm, businge:2012, businge:2013, businge:2015, dagenais:2008, mastrangelo:2015ACM}. For example, clients may decide to use internal APIs to improve performance, as a workaround for bugs, or to benefit from undocumented features. This usage is only possible because internal APIs are public, as the official and documented ones; and their usage is not checked by the Java compiler. To tackle this problem, a new module system is being proposed to Java, which will allow developers to explicitly declare the module elements they want to make available to external clients.\footnote{\url{http://openjdk.java.net/projects/jigsaw}} The Java compiler will use these declarations to properly encapsulate and check the usage of internal APIs. Therefore, our study reinforces the importance of introducing this new module system in Java, since we confirmed that changes in internal API elements are frequent. We also confirmed that API developers use the \mcode{public} keyword in Java with two distinct semantics (``public only to my code'' {\em vs} ``public to any code, including clients'').
\vspace{0.1cm}

\noindent{\em Tool Builders:} \apidiff\ is an useful tool both to API developers and clients. API developers can use the tool to document changes in their APIs, e.g.,~to automatically generate changelogs or release notes. API clients can also rely on \apidiff\ to  produce these documents, in order to better assess the effort to migrate to API versions that are not properly documented.
However, we also faced an important limitation when dealing with the output produced by \apidiff. Currently, {\sc move/rename} operations are detected as a removal ({\sc remove}) followed by an addition ({\sc add}) of an API element. As described in Section~\ref{section:apiDiff}, to generate the correct names for these operations, we had to manually inspect the output produced by \apidiff\ and the textual diff of the respective commits. Thus, we consider that \apidiff\ implementation can follow existing approaches and tools~\cite{sunghun:2005:functions_change, danilo:msr2017, fse2016-why-we-refactor} and  automatically detect the cases where {\sc remove} followed by an {\sc add} is indeed a {\sc rename} (when confined to the same class) or a {\sc move} (when involving different classes) refactoring.
\vspace{0.1cm}

\noindent{\em Researchers:} Although based on a limited number of 59 BCs, our study reveals opportunities to improve the state-of-the-art on API design, evolution, and analysis. \underline{First}, 
the study suggests that BCs are often motivated by the implementation of new features and that refactorings are usually performed at that moment, to support the implementation of the new code. In fact, a recent study on refactoring practices considering all types of GitHub projects, i.e.,~not restricted to libraries and frameworks, also shows that refactoring is mainly driven by the implementation of new requirements~\cite{fse2016-why-we-refactor}. 
Therefore, we envision a new research line on techniques and tools to recommend refactorings and related program redesign operations, when a new version of an API is under design. In other words, the focus should be on API-specific remodularization techniques, instead of global remodularization approaches, as commonly proposed in the literature~\cite{mitchell-automaticModularization-2006,praditwong-ModuleClustering-2011,abdeen-PackageCoupling-2009, anquetil-ExperimentsClustering-1999,spe2015}. \underline{Second}, the study suggests that BCs
are also motivated by a desire to reduce the number of API elements or reduce the possible usages of some elements (e.g.,~by making them \mcode{final}). Therefore, we envision research on API-specific static analysis tools (or API-specific linter tools), which could for example recommend the removal of useless parameters in API methods (as we found in 2 BCs), the insertion of a \mcode{final} modifier (as we found in 6 BCs) or even the removal of underused methods and classes (as we found in 2 BCs). The benefit in this case would be the recommendation of these changes at design time or during early usage phases, before the affected API elements gain clients and the change costs and impact increase. 
\underline{Third}, the answers of the third survey question suggest that BCs may have a minor impact on clients (but according to a small sample of six developers).
Thus, we envision further research on migration tools, which could help API clients to move to new API versions by providing recommendations on how to deal with trivial BCs~\cite{dagenais:2008, nguyen:2010, Wu10, zhang-AutomaticParameter-2012}. \underline{Fourth}, the answers of the last survey question show that some API developers can be reluctant to use the deprecation mechanism provided by Java. 
Essentially, they argue that deprecation increases maintenance burden, by requiring updates on multiple versions of the same API element.
Therefore, we also envision research on new and possibly lightweight mechanisms to API versioning. It is also possible to recommend the traditional mechanism only in special cases, particularly when the BCs might impact a large number of clients or require complex changes. 
\vspace{0.1cm}

\noindent {\em Practitioners:} The study also provides actionable 
results and guidelines to practitioners, especially to API developers. \underline{First}, we detected many unconfirmed BCCs in packages that do not have the terms \mcode{internal} or \mcode{experimental} (or similar ones) in their names. We recommend the usage of these names to highlight to clients the risks of using internal and unstable APIs.
\underline{Second}, the study also reveals that some BCs are caused by trivial programming mistakes, e.g.,~parameters that are never used. Since APIs are the external communication ports of libraries and frameworks, it is important that they are carefully designed and implemented. \underline{Third}, most BCs detected in the study require trivial changes in clients, at least according to six surveyed developers. Thus, API developers should 
carefully evaluate the introduction of BCs demanding complex migration efforts, which can trigger a strong rejection by clients. \underline{Fourth}, we listed good practices used by developers to document BCs, for example, changelogs and release notes.


\section{Threats to Validity}
\label{section:threatsValidity}

\noindent\emph{External Validity.} As usual in empirical software engineering, our findings are restricted to the studied subjects and cannot be generalized to other scenarios.
Nevertheless, we daily monitored a large dataset of 400  Java libraries and frameworks, during a period of 116 days (almost 4 months).
During this time, we questioned 102 developers about the motivations of breaking changes right after they had been performed.
Due to such numbers, we consider that our findings are based on representative libraries, which were assessed during a large period of time, with answers provided by developers while the subject was still fresh in their minds.
Moreover, our analysis is restricted to syntactical breaking changes, which result on compilation errors in clients.
BCs that modify the API behavior without changing its signature, usually named Behavioral Backward Incompatibilities~\cite{mostafa2017experience}, are outside of the scope of this paper. 
\vspace{0.1cm}

\noindent\emph{Internal Validity.} First, we use \apidiff\ to detect breaking changes between two versions of a Java library.
Although this tool was implemented and used in our previous research~\cite{laerte17}, an error on its result would introduce false positives in our analysis.
To mitigate this threat, we considered the breaking changes provided by the tool as \emph{candidates} and only assessed those confirmed by their developers, which represents 39\% of BCCs (see Sections~\ref{section:results-unconfirmed-bcs}).
Second, we reinforce the subjective nature of this study and its results. 
As discussed in Section~\ref{section:studyDesing-contacting-developers}, a thematic analysis was performed to elicit the reasons that drive API developers to introduce BCs.
Although this process was rigorously followed by two authors of the paper, the replication of this activity may lead to a different set of reasons.
To alleviate this threat, special attention was paid during the sequence of meetings held to resolve conflicts and to assign the final themes.
Third, against our belief, the trustworthiness and correctness of the responses is also a threat to be reported.
To  mitigate it, we strictly sent emails in no more than few days after the commits. 
This was important to guarantee a higher response rate and  reliable answers, once the modifications were still fresh on developers' minds.
\vspace{0.1cm}

\noindent\emph{Construct Validity.} The first threat relates to the selection of the Java libraries.
As discussed in Section~\ref{section:studyDesing-selection-libraries}, we automatically discarded, from the top-2,000 most popular Java projects on GitHub, the ones that do not have the following keywords in their short description: \emph{library(ies), API(s), framework(s)}.
Next, we manually discarded those that, although containing such words, do not actually represent a library.
Since this process is conservative in providing a reliable dataset of projects that are libraries, we can not guarantee that we retrieved the whole set of actual libraries from the 2,000 projects.
Second, our results stand on the agreement of developers on the detected BCCs.
As observed in Section~\ref{section:results-unconfirmed-bcs}, most developers pointed out that the detected changes refer to internal or low-level APIs, mentioning that it is unlikely that they could break clients.
However, previous research has shown that occasionally internal APIs are used by external clients~\cite{andre:fse2016, boulanger-ManagingConcern-2006-icsm, businge:2012, businge:2013, businge:2015, dagenais:2008, mastrangelo:2015ACM}.
Therefore, we might have excluded BCCs that could  actually impact clients, but we decided to follow the conservative decision of only considering BCCs perceived by developers as having a high potential to break existing clients.

\section{Related Work}
\label{section:relatedWork}

We organized related work in three subsections: (a) studies about breaking changes in APIs; (b) field studies using the firehouse interview method; (c) other studies on API evolution.

\subsection{Studies on Breaking Changes}

In a previous short paper, we report a preliminary study to reveal the reasons of API breaking changes in Java~\cite{laerte:sanerEra2016}. In this first study, we also use \apidiff\ to detect breaking changes. We contacted the principal developers of 49 libraries, asking them about the reasons of all breaking changes detected by \apidiff\ in previous releases of these libraries. By contrast, in this new study we contacted the precise developers responsible by a breaking change, right after it was introduced in the code; and we asked them to reveal the reasons for this specific breaking change. Furthermore, to identify breaking changes, we monitored all commits of a list of 400 Java libraries, during 116 days. As a consequence of the distinct  methodologies, in the first study we received valid answers of only seven developers (while in the present study we received 56 answers). From these seven answers, we extracted five reasons for breaking changes: API Simplification, Refactoring, Bug Fix, Dependency Changes, and Project Policy. The first four are also detected in the present study. However, the major reason for breaking changes reported in the present study (New Feature) was not detected in the preliminary one. 

In another related study~\cite{laerte17}, we investigate breaking changes in 317 real-world Java libraries, including 9K releases and
260K client applications. We show that 15\% of the API changes
break compatibility with previous versions and that the frequency of
breaking changes increases over time. Using data from the BOA ultra-large dataset~\cite{boa:2013}, we report that less than 3\% of the breaking changes impact clients. To reach this result, we considered all breaking changes detected by \apidiff. However, in the present paper, we found that only 39\% of the BCCs are viewed by developers as having a major potential to break existing clients.

Dig and Johnson~\cite{johnson:acm2006} studied API changes in five frameworks and libraries (Eclipse, Mortgage, Struts, Log4J, and JHotDraw). They report that more than 80\% of the
breaking changes in these systems were due to refactorings. By contrast, using a large dataset of 400 popular Java libraries and frameworks, we also found that BCs are usually related to refactorings, but at a lower rate (47\%). Moreover, we listed two other important motivations for breaking changes: to support the implementation of new features and to simplify and reduce the number of API elements. Bogart \textit{et al.}~\cite{thung16} conducted a study to understand how developers plan, negotiate, and manage breaking changes in three software ecosystems: Eclipse, R/CRAN, and Node.js/npm. After interviewing key developers in each ecosystem, they report that a core value of the Eclipse community is long-term stability; therefore, breaking changes are rare in Eclipse. R/CRAN values snapshot consistency, i.e.,~the newest version of every package should be always compatible with the newest version of every other package in the ecosystem. Once snapshot consistency is preserved, breaking changes are not a major concern in R/CRAN. Finally, breaking changes in Node.js/npm  are viewed  as necessary for progress and innovation. In the interviews, the participants also mentioned three general reasons for breaking changes: technical debt (i.e.,~to improve maintainability), to fix bugs,  and to improve performance. The first two motivations appear in our study, but we did not detect breaking changes motivated by performance improvements. However, these answers should be interpreted as general reasons for breaking changes, as perceived by the interviewed developers. By contrast, in our study the goal was to reveal reasons for specific breaking changes, as declared by developers right after introducing them in the source code of popular Java libraries and frameworks.

\subsection{Studies using Firehouse Interviews}

A {\em firehouse interview} is one that is conducted right after the event of interest has happened~\cite{diffusionInnovations-2003}. The term relates to the difficulty of performing qualitative studies about unpredictable events, like a fire. In such cases, researchers should act like firemen after an alarm; they should rush to the firehouse, instead of waiting the event to be concluded to start their research. 
In our study, the events of interest are API breaking changes; and firehouse interviews allowed us to collect the reasons for theses changes right after they were committed to GitHub repositories. In software engineering research, firehouse interviews were previously used to investigate bugs just fixed by developers~\cite{murphyHill-designBugFixes-2013, murphyHill-SpaceBugFixes-2015}, but using face-to-face interviews with eight Microsoft engineers. Silva \textit{et al.}~\cite{fse2016-why-we-refactor} were the first to use firehouse interviews to contact GitHub developers by email. Their goal was to reveal the reasons behind refactorings applied by these developers; in this case, they also used a tool to automatically detect refactorings performed in recent commits. They sent e-mails to 465 developers and received 195 answers (42\% of response ratio). Mazinanian \textit{et al.}~\cite{mazinanian:2017} used a similar approach, but to understand the reasons why developers introduce lambda expressions in Java. They sent emails to 351 developers and received 97 answers (28\% of response ratio). In our study, we contacted 102 developers and received 56 answers (55\% of response ratio).

\subsection{Studies on API Evolution}

Several studies have been proposed to support API evolution and client developers.
Chow and Notkin~\cite{King96} present an approach where library developers themselves annotate the changed methods with replacement rules.
Henkel and Diwan~\cite{henkel:2005} propose a tool that captures and replays API evolution refactorings.
Kim~\textit{et al.}~\cite{Kim09a} support computing differences between two versions of a system. 
Nguyen~\textit{et al.}~\cite{nguyen:2010} use graph-based techniques to help developers migrate from one library version to another.
Other studies focus on extracting API evolution rules from source code.
For example, Schafer~\textit{et al.}~\cite{Scha08} mine library change rules from client systems, while Dagenais and Robillard~\cite{dagenais:2008} suggest API replacements based on how libraries adapt to their own changes.
Also in this context, Meng~\textit{et al.}~\cite{Meng12} propose a history-based matching approach to support API evolution. 

In a large-scale study, Robbes~\textit{et al.}~\cite{Robb12} assess the impact of API deprecation in a Smalltalk ecosystem.
Recently, the authors also evaluated the impact in the context of the Java programming language~\cite{sawant2016, sawant:ese2017}.
In this study, they found that some API deprecation have large impact on the ecosystem under analysis and that the quality of deprecation messages should be improved.
Jezek \textit{et al.}~\cite {jezek:2015} study 109 Java open-source programs and 564 program versions, showing that APIs are commonly unstable. 
Raemaekers \textit{et al.} \cite{raemaekers12} investigate API stability  with the support of four proposed metrics, based on method removal and implementation change.
In the context of mobile development, McDonnell \textit{et al.}~\cite{McDo13} investigate stability and adoption of the Android API. 
In this study, the authors show that APIs are updated on average 115 times per month, representing a rate faster than clients' update. 

Some studies investigate the usage and evolution of internal APIs, i.e., public but unstable and undocumented APIs that should not be used by client applications~\cite{businge:2012, businge:2013, businge:2015, mastrangelo:2015ACM, andre:fse2016}.
In this context, Businge~\textit{et al.}~\cite{businge:2012} study the survival of Eclipse plugins, and classify them in two categories: plugins depending on internal APIs and plugins depending only on official APIs.
In an extended study~\cite{businge:2015}, the authors present that 44\% of $512$ Eclipse plugins depend on internal APIs.
In addition, the same authors investigate the reasons why developers do use internal APIs~\cite{businge:2013}.
For example, they detect cases where developers do not read documentation (so they are not aware of the risks), but also cases where developers deliberately use internal APIs to benefit from advanced features, not available in the official APIs.
Mastrangelo~\textit{et al.}~\cite{mastrangelo:2015ACM} show that clients commonly use the internal API \mcode{sun.misc.Unsafe} provided by JDK.
Recently, Hora~\textit{et al.}~\cite{andre:fse2016} studied the transition of internal APIs to public ones, aiming to support library developers to deliver better API modularization.The authors also performed a large analysis to assess the usage of internal APIs.
In our survey, several developers mentioned that the breaking changes happened in public but internal or low-level APIs that clients should not rely on.
Notice, however, that the related literature points in the opposite direction: client developers tend to use internal APIs.

\section{Conclusion}
\label{section:conclusion}


Libraries and frameworks are key instruments to promote reuse an increase productivity in modern software development. Ideally, software libraries and frameworks should provide stable and backward-compatible APIs to their clients. However, the practice reveals that breaking changes (BCs) are common. In this paper, we described a large-scale empirical study (400 libraries, 4-month long period, 282 possible breaking changes, 56 developers contacted by email) to understand {\em why} and {\em how} developers break APIs in Java. 
By using a firehouse interview method,
we found that BCs are mainly motivated by the implementation of new features, to simplify the number of API elements, and to improve maintainability. 
The most common BCs are due to refactorings (47\%); regarding the programming elements affected by BCs, most are methods (59\%). According to the surveyed developers, the effort on clients to migrate to new API versions, after BCs, is minor.
We also listed some strategies to document BCs, like release notes and changelogs. Last but not least, we presented an extensive list of empirically-justified implications of our study, targeting four distinct audiences: programming languages designers, tool builders, software engineering researchers, and API developers. However, such implications should be viewed and interpreted with care, since they are derived from considering only 59 BCs and a single programming language (Java).

Further studies can consider other software ecosystems and programming languages (particularly, dynamic languages); other research methodologies (e.g.,~semi-structured interviews); and also provide a quantitative and qualitative assessment of the impact of breaking changes in the other protagonists of this story: the developers who depend on APIs affected by breaking changes.

\section*{Acknowledgments}

\noindent We thank the 56 GitHub developers who participated in our study and shared their ideas and practices about breaking changes. This research is supported by grants from FAPEMIG (process CEX-PPM-00490-17) and CNPq (process 306554/2015-1).

\bibliographystyle{ieeetr}
\bibliography{bib}

\end{document}